\documentclass[aps,prd,twocolumn,showpacs,preprintnumbers,amsmath,amssymb]{revtex4}
\usepackage{graphicx}
\usepackage{dcolumn}
\usepackage{bm}
\usepackage{amssymb,amsmath}
\usepackage{latexsym}

\allowdisplaybreaks

\begin{document}

\title{Optical position meters analyzed in the non-inertial reference frames}
\author{Sergey P. Tarabrin and Alexander A. Seleznyov}
\affiliation{Faculty of Physics, Moscow State University, Moscow,
119992, Russia} \email{tarabrin@phys.msu.ru}
\date{\today}

\begin{abstract}
In the framework of General Relativity we develop a method for
analysis of the operation of the optical position meters in their
photodetectors proper reference frames. These frames are
non-inertial in general due to the action of external fluctuative
forces on meters test masses, including detectors. For comparison we
also perform the calculations in the laboratory (globally inertial)
reference frame and demonstrate that for certain optical schemes
laboratory-based analysis results in unmeasurable quantities, in
contrast to the detector-based analysis. We also calculate the
response of the simplest optical meters to weak plane gravitational
waves and fluctuative motions of their test masses. It is
demonstrated that for the round-trip meter analysis in both the
transverse-traceless (TT) and local Lorentz (LL) gauges produces
equal results, while for the forward-trip meter corresponding
results differ in accordance with different physical assumptions
(e.g. procedure of clocks synchronization) implicitly underlying the
construction of the TT and LL gauges.
\end{abstract}
\preprint{LIGO-P080043} 

\maketitle

\section{Introduction}\label{sec_intro}
Optical position meters can be thought of as constituent parts of
laser interferometers such as Michelson or Mach-Zehnder. For
instance, long-baseline Michelson interferometer is the composition
of two round-trip position meters: two optical waves, emerging from
and returning to approximately the same spacial point, inside
interferometer arms carry the information about displacements of the
end-mirrors. Upon their arrival to photodetector the latter measures
the relative phase of two waves proportional to the relative
displacement of the end-mirrors.

Currently long-baseline optical interferometers, such as LIGO
\cite{website_LIGO}, are the world most sensitive instruments for
measuring the relative displacements between the test masses, which
might be produced by the gravitational waves (GWs) coming to the
Earth from astrophysical sources. Traditionally in literature
interaction of the gravitational waves with laser interferometers is
considered in the framework of the so-called transverse-traceless
(TT) gauge
\cite{1973_gravitation,2003_clas_phys_GW,2005_basics_of_gw_theory}.
Though the TT-based analysis of the GW detectors is usually
exceptionally simple, it is sometimes hard to be justified
physically since the major requirement of the TT gauge is the ideal
inertiality of the test masses. In other words, in order to validate
the TT-based analysis one must require that the test masses strictly
follow the geodesics of the GW space-time, i.e. no non-gravitational
forces are present. However, at least in the Earth-bound
experimental installations test masses undergo various non-geodesic
motions due to noises in the environment (such as seismic noise), in
the test masses themselves (thermal noise), due to measurement
devices (back-action noise) and others.

In order to deal correctly with external forces when analyzing the
operation of an interferometer (or optical position meter) one
should perform the calculations in the proper reference frame of its
detector since it is the device that produces an experimentally
observable quantity. Since detector is subjected to the action of
fluctuative forces in general, its proper reference frame is
non-inertial. This is particularly important for the certain class
of the GW detectors called displacement-noise-free interferometers,
which are mostly of Mach-Zehnder
\cite{2004_DNF_GW_detection,2006_DTNF_GW_detection,
2006_interferometers_DNF_GW_detection,2007_time_delay} or
Fabry-Perot \cite{DFI_FP} type, where non-geodesic motion of
detector(s) may significantly affect the operation of an
interferometer limiting its sensitivity to GWs. The main goal of
this paper is to develop the method of solving certain
electrodynamical problems in such non-inertial reference frames.

\section{Space-time in the vicinity of an accelerated observer}
\subsection{Metric tensor}\label{sec_metric_tensor}
We start from the introduction of space-time metric in the vicinity
of an accelerated observer. First, consider the laboratory (globally
inertial) frame with Minkowski metric
$ds^2=\eta_{\alpha\beta}dx^{\alpha}dx^{\beta}$, where
$\eta_{\alpha\beta}=\textrm{diag}(-1,1,1,1)$. Greek indices run over
$0,1,2,3$.

It is known that the non-relativistic ($v^2/c^2\ll1$) coordinate
transformation \cite{2003_clas_phys_acc_obs}
\begin{equation}
    x=x'+\frac{a_x(t')t'^2}{2},\quad
    t=t'\left[1+\frac{a_x(t')x'}{c^2}\right].\label{coord trans}
\end{equation}
brings us from the laboratory frame to the proper reference frame of
the observer (reference body for definiteness) moving with
acceleration $a_x(t')$ along the $x$-axis. Throughout the paper
prime denotes the physical quantity evaluated in the observer's
proper reference frame. According to the transformation law, metric
in the vicinity of an accelerated observer, accurate to linear order
of $x'$, takes the form:
\begin{align}
    ds^2&=g_{\alpha\beta}(x'^{\mu})dx'^\alpha dx'^\beta\nonumber\\
    &=-(c\,dt')^2\left[1+\frac{2}{c^2}\,a_x(t')x'\right]+dx'^2+dy'^2+dz'^2.
    \label{interval}
\end{align}

We assume that the acceleration $a_x(t')$ is so small that for all
$x'$ and $t'$ condition $|2a_x(t')x'/c^2|\ll1$ is fulfilled. In
particular, for the ground-based interferometers one can estimate
$a_x\lesssim\Omega_{\max}^2\xi$ with $\Omega_{\max}\sim10^3$ Hz (the
upper boundary of the operating frequency band), $\xi\sim 10^{-19}$
m (typical values of the fluctuative displacements), $x\sim 10^3$ m
(typical interferometer arm length), so that
$|2a_xx'/c^2|\sim10^{-24}$. Therefore, we will use the methods of
linearized theory in full similarity with the theory of linear
gravitational waves; we will keep only the 0th and the 1st order in
$a_x(t')$ terms further.

Metric tensor corresponding to interval (\ref{interval}) and the
determinant of its matrix are:
\begin{align}
    g_{\alpha\beta}&=
        \begin{pmatrix}
            -1-2a_x(t')x'/c^2&0&0&0\\
            0&1&0&0\\
            0&0&1&0\\
            0&0&0&1\\
        \end{pmatrix},\label{metric}\\
        g&=\det(g_{\alpha\beta})=-1-\frac{2}{c^2}\,a_x(t')x'.
        \nonumber
\end{align}

\subsection{Test masses equation of motion}\label{sec_eq_mot}
Since we consider the motion of the test masses along the $x'$-axis,
the only Christoffel symbols needed are $\Gamma^{\mu}_{\nu\lambda}$
with $\mu=1$: $\Gamma_{00}^{1}=a_x(t')/c^2$, $\Gamma_{01}^{1}=0$.
Geodesic equation for the $x$-axis reads:
\begin{equation*}
    \frac{d^2x'^1}{ds^2}+\Gamma_{00}^{1}\left(\frac{dx'^0}{ds}\right)^2+
    2\Gamma_{01}^{1}\,\frac{dx'^0}{ds}\frac{dx'^1}{ds}+
    \Gamma_{11}^{1}\left(\frac{dx'^1}{ds}\right)^2=0.
\end{equation*}
In the non-relativistic approximation $ds\approx c\,dt'$ and
$(dx'^1/ds)^2\approx(v'/c)^2\ll1$, therefore we obtain the following
equation of motion (for strict derivation see Ref.
\cite{1978_inert_grav_eff}):
\begin{equation*}
    \frac{d^2x'}{dt'^2}=-a_x(t'),
\end{equation*}
which coincides exactly with the Newtonian law of motion in the
non-inertial frame. If the test mass $m$ is also subjected to some
external force $F_x(t')$ as seen from the laboratory frame, then the
latter should be added to the right side of the equation:
\begin{equation}
    \frac{d^2x'}{dt'^2}=-a_x(t')+\frac{F_x(t')}{m}.
    \label{eq_of_mot}
\end{equation}
Thus, in the absence of the observer's acceleration (when $x=x'$ and
$t=t'$) we obtain the Newtonian motion law of the test mass in the
laboratory frame: $d^2x/dt^2=F_x(t)/m$.

For simplicity we assume that the test mass and the observer stay in
rest, separated by a distance $x'_0=x_0=\textrm{const}$, with
respect to the laboratory frame in the absence of all forces in
order not to consider the effects of uniform motion.

Below we will consider the problems where test masses undergo tiny
fluctuative displacements under the influence of external forces,
i.e. $F_x/m$ is of the same order of smallness as $a_x$. Thus, Eq.
(\ref{eq_of_mot}) allows significant simplification: according to
the transformation law (\ref{coord trans}) coordinate time in the
observer's frame $t'$ and the one in the laboratory frame $t$ differ
in the amount proportional to $a_xx'/c^2$ which is the quantity of
the 1st order of smallness. Therefore, up to the 1st order
$a_x(t')=a_x(t)$ and $F_x(t')=F_x(t)$. Under the listed assumptions
Eq. (\ref{eq_of_mot}) can be integrated in the following form:
\begin{equation}
    x'(t)=x_0+\int_{-\infty}^{t}dt_1\int_{-\infty}^{t_1}
    dt_2\left[-a_x(t_2)+\frac{F_x(t_2)}{m}\right].
    \label{law_of_mot}
\end{equation}
It will be convenient to separate the 0th order and the 1st order
summands: $x'(t)=x_0+\delta x'(t)$, $|\delta x'|\ll |x_0|$,
\begin{equation}
    \delta x'(t)=\xi(t)-\xi_{\textrm{ref}}(t),
    \label{1st_order_law_of_mot}
\end{equation}
where $\xi(t)$ is the result of double integration of $F_x(t)/m$ and
$\xi_{\textrm{ref}}(t)$ is the result of double integration of
$a_x(t)$. The physical meaning of these quantities is clear: $\xi$
and $\xi_{\textrm{ref}}$ are the displacements of the test mass and
the reference mass correspondingly with respect to the laboratory
frame, while $\delta x'$ is the displacement of the test mass with
respect to the reference mass (i.e. with respect to the proper
reference frame of the reference mass).

\section{Electromagnetic wave in the space-time of an accelerated
observer}In the interferometric experiments an observer studies the
motion of the test masses by sending and receiving the reflected
light waves. According to the equivalence principle an accelerated
frame (of the observer) is equivalent to some gravitational field
which is known to impose the distributed redshift on electromagnetic
waves. Thus it is necessary to calculate the propagation of
electromagnetic waves in the space-time of an accelerated observer
in order to obtain a complete description of an interferometer.

\subsection{Wave equation}\label{sec_wave_eq}
We will derive the wave equation from the second pair of Maxwell's
equations without the sources:
\begin{equation*}
    \frac{1}{\sqrt{-g}}\frac{\partial}{\partial x'^\beta}
    \left(\sqrt{-g}F'^{\alpha\beta}\right)=0.
\end{equation*}
Here $F'_{\mu\nu}=\partial'_\mu A'_\nu-\partial'_\nu A'_\mu$ and
$A'^{\mu}=(A'^0,A'^1,A'^2,A'^3)$ is the 4-potential of
electromagnetic field. Substituting the definition of $F'_{\mu\nu}$
into the field equations we obtain:
\begin{multline*}
    \left(\partial'_\beta\sqrt{-g}\right)\left(\partial'^\alpha A'^\beta-
    \partial'^\beta A'^\alpha\right)\\+
    \sqrt{-g}\left(\partial'^\alpha\partial'_\beta A'^\beta-
    \partial'_\beta\partial'^\beta A'^\alpha\right)=0.
\end{multline*}
Let us impose Lorentz or Coulomb gauge (this will influence only the
procedure of quantization) so that $\partial'_\beta A'^\beta$
vanishes. Remind now that the non-zero components of metric tensor
(\ref{metric}) are $g_{00},\ g_{11},\ g_{22}$ and $g_{33}$ and let
us assume that the vector-potential describes the propagation along
the $x'$-axis of the plane electromagnetic wave polarized along the
$z'$-axis, i.e. $A'^\alpha=(0,0,0,A')$ and
$A'=A'(x'^0,x'^1)=A'(x',t')$. Remind also that $g=g(x',t')$.
Therefore, wave equation reduces to:
\begin{multline*}
    \left(\partial'_0\sqrt{-g}\right)\left(-g^{00}\partial'_0A'\right)+
    \left(\partial'_1\sqrt{-g}\right)\left(-g^{11}\partial'_1A'\right)\\-
    \sqrt{-g}\left(g^{00}\partial'_0\partial'_0A'+
    g^{11}\partial'_1\partial'_1A'\right)=0.
\end{multline*}
Substituting here the components of metric tensor in an explicit
form and linearizing the equation with respect to the terms
containing $a_x(t')$, we finally obtain the following scalar wave
equation:
\begin{align}
    \frac{1}{c^2}\frac{\partial^2A'}{\partial t'^2}-\frac{\partial^2A'}{\partial x'^2}&=
    \frac{a_xx'}{c^2}\left(\frac{1}{c^2}\frac{\partial^2A'}{\partial t'^2}+
    \frac{\partial^2A'}{\partial x'^2}\right)\nonumber\\&\quad-
    \frac{\dot{a}_xx'}{c^3}\,\frac{1}{c}\frac{\partial A'}{\partial t'}+
    \frac{a_x}{c^2}\,\frac{\partial A'}{\partial x'}.
    \label{wave_eq}
\end{align}
Here $\dot{a}_x=da_x(t')/dt'$. The right side of this equation
describes the redshift produced by the non-inertiality of the
reference frame.

\subsection{Solution of the wave equation}\label{sec_wave_eq_solution}
It is convenient to solve the obtained equation using the method of
successive approximations in full similarity with the solution of
wave equation in Ref \cite{2007_GW_FP_LL}. We shall keep only the
0th and the 1st order in $a_x(t')$ terms:
$A'(x',t')=A'^{(0)}(x',t')+A'^{(1)}(x',t'), \
|A'^{(1)}|\sim|(a_xx'/c^2)A'^{(0)}|\ll |A'^{(0)}|$. The 0th order
corresponds to the unaccelerated observer which stays in rest in the
laboratory frame. Thus, solution of the 0th order can be represented
as a sum of plane monochromatic waves traveling in positive and
negative directions of the $x'$-axis with amplitudes and frequency
measured in the laboratory frame in the state of rest. We denote
``positive'' wave with index '+' and ``negative'' wave with index
'--':
\begin{align}
    &A'^{(0)}(x',t')=A_+'^{(0)}(x',t')+A_-'^{(0)}(x',t'),\label{0th_order_solution}\\
    &A_\pm'^{(0)}=A_{\pm0}e^{-i(\omega_0t'\mp k_0x')}+\textrm{c.c.},\nonumber
\end{align}
where $k_0=\omega_0/c$. Evidently, in the 0th order $t'=t$ and
$x'=x$. Amplitudes and frequency are derived from some initial and
boundary problems and we shall keep them undefined until next
section. The 1st order equation is:
\begin{align}
    \frac{1}{c^2}\frac{\partial^2A_\pm'^{(1)}}{\partial t'^2}-
    \frac{\partial^2A_\pm'^{(1)}}{\partial x'^2}&=
    \frac{a_xx'}{c^2}\left(\frac{1}{c^2}\frac{\partial^2A_\pm'^{(0)}}{\partial t'^2}+
    \frac{\partial^2A_\pm'^{(0)}}{\partial x'^2}\right)\nonumber\\&\quad-
    \frac{\dot{a}_xx'}{c^3}\,\frac{1}{c}\frac{\partial A_\pm'^{(0)}}{\partial t'}+
    \frac{a_x}{c^2}\,\frac{\partial A_\pm'^{(0)}}{\partial x'}.
    \label{1st_order_wave_eq}
\end{align}
The general solution of this equation can be represented as a sum of
``positive'' and ``negative'' waves:
\begin{equation}
    A'^{(1)}(x',t')=A_+'^{(1)}(x',t')+A_-'^{(1)}(x',t').
    \label{sum}
\end{equation}
Clearly, they can be treated independently. Remind, that
$g_{00}(0,t)=-1$ and therefore, we must demand that
\begin{equation}
    A_+'^{(1)}(0,t')=A_-'^{(1)}(0,t')=0.
    \label{initial_conditions}
\end{equation}
Physically these initial conditions mean that both the light waves
$A'_{\pm}(x',t')$ experience no redshift at $x'=0$, i.e. the
solution of full Eq. (\ref{wave_eq}) at $x'=0$ is
$A'_\pm(0,t')=A'^{(0)}_\pm(0,t')=A_{\pm0}e^{-i\omega_0t}+A_{\pm0}^*e^{i\omega_0t}$
(remind, that $t'=t$ at $x'=0$ according to Eqs. (\ref{coord
trans})). The solution of Cauchy problem (\ref{1st_order_wave_eq} --
\ref{initial_conditions}) is obtained in Appendix
\ref{sec_solution_wave_eq}. For slow enough mechanical motions (as
compared to the optical frequency) we have:
\begin{align}
    &A'^{(1)}_\pm(x',t')=A_{\pm0}w'_\pm(x',t')e^{-i(\omega_0t'\mp k_0x')}
    +{\textrm{c.c.}},\label{1st_order_solution}\\
    &w'_\pm(x',t')=-ik_0\dot{\xi}_{\textrm{ref}}(t')\,\frac{x'}{c}\pm
    ik_0\Bigl[\xi_{\textrm{ref}}(t')-\xi_{\textrm{ref}}(t'\mp x'/c)\Bigr].\nonumber
\end{align}
Remind, that $\xi_{\textrm{ref}}(t')$ is the result of integration
of $a_x(t')$ such that $a_x(t')=\ddot{\xi}_{\textrm{ref}}(t')$.
Throughout the paper below we will omit the ``c.c.'' notation for
briefness.

Similarly to the previous section we may replace $t'\rightarrow t$
and $x'\rightarrow x$ in $w'(x',t')$ without introducing an error in
the 1st order:
\begin{equation*}
    w'_\pm(x,t)=-ik_0\dot{\xi}_{\textrm{ref}}(t)\,\frac{x}{c}\pm
    ik_0\Bigl[\xi_{\textrm{ref}}(t)-\xi_{\textrm{ref}}(t\mp x/c)\Bigr].
\end{equation*}
Whenever it is convenient we will use either $w'_\pm(x',t')$ or
$w'_\pm(x,t)$ below.

Several features of $w'_\pm(x,t)$ are worth noting. First, according
to the physical sense $w_\pm(0,t)=w_\pm(0,t')=0$, i.e. the frequency
of electromagnetic wave is not redshifted in the immediate vicinity
of an observer. Second, for small enough $x/c$ (in spectral domain
this corresponds to $\Omega x/c\ll1$ limit) $w'_\pm(x,t)$ has the
$O[k_0(x/c)^2a_x(t)]$ asymptotic. This agrees with the relativity
principle: electromagnetic wave senses only the acceleration of the
reference frame, not the displacement $\xi_{\textrm{ref}}$ or
velocity $\dot{\xi}_{\textrm{ref}}$. Finally, since
$\xi_{\textrm{ref}}(t)$ and $\dot{\xi}_{\textrm{ref}}(t)$ are the
pure real quantities, $w'_\pm(x,t)$ are the pure imaginary
quantities and therefore describe the influence of the acceleration
on the phase (amplitude is affected beginning from the 2nd order):
\begin{align*}
    A'_\pm(x',t')&=A_{\pm0}\Bigl[1+w'_\pm(x',t')\Bigr]e^{-i(\omega_0t'\mp k_0x')}\nonumber\\
    &\approx A_{\pm0}e^{-i(\omega_0t'\mp k_0x')+w'_\pm(x',t')}.
\end{align*}

\section{Round-trip position meter}\label{sec_round_trip_detector}
\begin{figure}[h]
\includegraphics[scale=0.6]{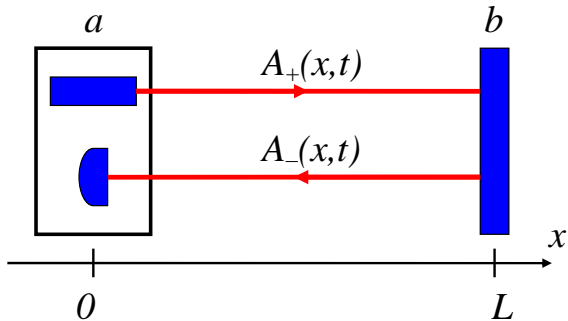}
\caption{A round-trip position meter. Laser mounted on test mass $a$
emits the wave which is reflected from the movable mirror $b$.
Detector mounted on test mass $a$ detects the reflected wave and
measures the acquired phase shift. In the state of rest the distance
between the test masses equals to $L$.}\label{pic_scheme}
\end{figure}

Let us consider now the optical scheme of the round-trip position
meter illustrated in Fig. \ref{pic_scheme}: light wave emitted by
the laser mounted on test mass $a$ reaches the absolutely reflective
mirror (test mass $b$) and is reflected back to the detector mounted
on test mass $a$. We are interested in the phase shift acquired by
the light wave. Such a position meter might be a constituent part (a
single arm) of a Michelson interferometer. In the state of rest the
distance between the test masses equals to $L$.

\subsection{Analysis in the inertial frame}
First we analyze the situation from the viewpoint of observer in the
laboratory (globally inertial) reference frame. Coordinates of the
test masses $a$ and $b$ are $x_a(t)=0+\xi_a(t)$ and
$x_b(t)=L+\xi_b(t)$ correspondingly. Remind that $|\xi_{a,b}|\ll L$.
We also approximate the light wave as noiseless (see Appendix
\ref{sec_optical_noise} for generalization). Thus we may write the
wave that laser emits in the following form:
\begin{equation}
    A_+(x,t)=A_{+0}\exp\left\{-i\omega_0\left[t-\frac{x-x_a(t-x/c)}{c}\right]\right\},
    \label{A+}
\end{equation}
The wave reflected from the mirror is described by vector-potential
\begin{equation}
    A_-(x,t)=A_{-0}e^{-i(\omega_0t+k_0x)}+a_-(x,t)e^{-i(\omega_0t+k_0x)},
    \label{A-}
\end{equation}
where $a_-(x,t)$, which carries the information about the acquired
phase shift, can be represented as the Fourier integral:
\begin{equation*}
    a_-(x,t)=\int_{-\infty}^{+\infty}a_-(\Omega+\omega_0)e^{-i\Omega(t+x/c)}\,
    \frac{d\Omega}{2\pi}.
\end{equation*}

To find the relationship between the incident and the reflected
waves we impose the boundary condition which states that the vector
potential vanishes on the mirror surface:
\begin{equation*}
    A_+(x_b(t),\,t)+A_-(x_b(t),\,t)=0.
\end{equation*}
Substituting fields (\ref{A+}) and (\ref{A-}) into this equation and
keeping only the 0th and the 1st order terms we obtain:
\begin{multline*}
    A_{+0}e^{ik_0L}\Bigl[1+ik_0\xi_b(t)-ik_0\xi_a(t-\tau)\Bigr]\\
    +A_{-0}e^{-ik_0L}\Bigl[1-ik_0\xi_b(t)\Bigr]+a_-(L,t)e^{-ik_0L}=0.
\end{multline*}
Here $\tau=L/c$. The 0th order solution is
\begin{equation*}
    A_{-0}=-A_{+0}e^{2i\omega_0\tau}.
\end{equation*}
In the 1st order we obtain:
\begin{equation*}
    a_-(L,t)=-A_{+0}e^{2i\omega_0\tau}ik_0\Bigl[2\xi_b(t)-\xi_a(t-\tau)\Bigr],
\end{equation*}
or
\begin{equation*}
    a_-(0,t)=
    -A_{+0}e^{2i\omega_0\tau}ik_0\Bigl[2\xi_b(t-\tau)-\xi_a(t-2\tau)\Bigr],
\end{equation*}
according to the wave-like representation of $a_-(x,t)$.

Detection of the reflected wave takes place at point
$x=x_a(t)=\xi_a(t)$. Therefore, total variation of the optical wave
$\delta a(t)$ per round trip equals to
$a_-(0,t)-A_{-0}ik_0\xi_a(t)$:
\begin{equation*}
    \delta a(t)=
    -A_{+0}e^{2i\omega_0\tau}ik_0\Bigl[2\xi_b(t-\tau)-\xi_a(t-2\tau)-\xi_a(t)\Bigr].
\end{equation*}
We are interested in the phase shift $\delta\Psi$ described by the
term in the square brackets:
\begin{equation*}
    \delta\Psi(t)=-k_0\Bigl[\xi_a(t)-2\xi_b(t-\tau)+\xi_a(t-2\tau)\Bigr].
\end{equation*}
Evidently, this phase shifts describes a round trip of light wave
with correct time delays. If the entire system moves as a rigid
body, i.e. $\xi_a(t)=\xi_b(t)$, then $\delta\Psi(t)\approx
-k_0\ddot{\xi}_a(t)\tau^2$. This result agrees with the relativity
principle: no absolute displacement $\xi_a$ or velocity
$\dot{\xi}_a$ can be measured.

\subsection{Analysis in the non-inertial
frame} Now we will consider the same situation in the proper
reference frame of the test mass $a$, where detector is mounted, and
compare the result with the one of the laboratory-frame analysis.

Since test mass $a$ is the reference body, its equation of motion is
$x'_a(t)=0$. The coordinate of test mass $b$ is $x'_b(t)=L+\delta
x'_b(t)$ with $|\delta x'_b|\ll L$.

In its proper reference frame laser emits the wave described by
vector-potential
\begin{equation*}
    A'_+(x',t')=A_{+0}\Bigl[1+w'_+(x',t')\Bigr]e^{-i(\omega_0t'-k_0x')},
\end{equation*}
according to the results of Sec. \ref{sec_wave_eq_solution}.
Reflected wave in the reference frame of test mass $a$ is described
by vector-potential
\begin{align*}
    A'_-(x',t')&=A_{-0}\Bigl[1+w'_-(x',t')\Bigr]e^{-i(\omega_0t'+k_0x')}\nonumber\\&\quad+
    a'_-(x',t')e^{-i(\omega_0t'+k_0x')}.
\end{align*}
Here $a'_-(x',t')$ has the same physical meaning as in the previous
section and thus has the 1st order of smallness.

Substituting both waves into the boundary condition
\begin{equation*}
    A'_+(x'_b(t'),\,t')+A_-(x'_b(t'),\,t')=0.
\end{equation*}
we obtain:
\begin{multline*}
    A_{+0}e^{ik_0L}\Bigl[1+ik_0\delta x'_b(t)+w'_+(L,t)\Bigr]\\
    +A_{-0}e^{-ik_0L}\Bigl[1-ik_0\delta x'_b(t)+w'_-(L,t)\Bigr]\\
    +a'_-(L,t)e^{-ik_0L}=0.
\end{multline*}
The 0th order solution is similar to the previous case:
$A_{-0}=-A_{+0}e^{2i\omega_0\tau}.$ In the 1st order we obtain:
\begin{multline*}
    a'_-(0,t)=-A_{+0}e^{2i\omega_0\tau}\\
    \times\Bigl[2ik_0\delta x'_b(t-\tau)+w'_+(L,t-\tau)-w'_-(L,t-\tau)\Bigr].
\end{multline*}

Since detection of the reflected wave takes place at point
$x'=x'_a(t)=0$ in the reference frame we work in, total variation of
the optical wave coincides with $a'_-(0,t)$. Phase shift describing
the round trip is:
\begin{align}
    &i\,\delta\Psi'(t)\nonumber\\
    &=\Bigl[2ik_0\delta x'_b(t-\tau)+w'_+(L,t-\tau)-w'_-(L,t-\tau)\Bigr]\nonumber\\
    &=2ik_0\delta x'_b(t-\tau)-ik_0\Bigl[\xi_a(t)-2\xi_a(t-\tau)+\xi_a(t-2\tau)\Bigr].
    \label{phase}
\end{align}
To analyze its physical meaning we first expand it in the series of
$\tau$ keeping all the terms up to $\ddot{\xi}_a\tau^2$:
\begin{equation*}
    \delta\Psi'(t)\approx 2k_0\delta x'_b(t)-k_0\ddot{\xi}_a(t)\tau^2.
\end{equation*}
If the entire system moves as a rigid body, i.e. $\delta x'_b(t)=0$,
then $\delta\Psi'(t)\approx -k_0\ddot{\xi}_a(t)\tau^2$ in full
agreement with the relativity principle.

Let us now substitute solution (\ref{1st_order_law_of_mot}) for
$\delta x'_b$ into the phase shift (\ref{phase}), keeping in mind
that $\xi_{\textrm{ref}}=\xi_a$ and $\xi=\xi_b$:
\begin{align*}
    \delta\Psi'(t)&=2k_0\Bigl[\xi_b(t-\tau)-\xi_a(t-\tau)\Bigr]\\
    &\quad-k_0\Bigl[\xi_a(t)-2\xi_a(t-\tau)+\xi_a(t-2\tau)\Bigr]\\
    &=-k_0\Bigl[\xi_a(t)-2\xi_b(t-\tau)+\xi_a(t-2\tau)\Bigr].
\end{align*}
Thus, the obtained phase shift $\delta\Psi'(t)$ coincides with the
phase shift $\delta\Psi(t)$ in the laboratory frame. In other words,
consideration in both the frames results in equal measurable
quantities. This coincidence owes to the fact that in our round-trip
scheme phases of both the emitted and reflected waves are measured
at the same spacial point; in the proper frame of test mass $a$ this
point is located at $x'=0$ where rate of the built-in-test-mass-$a$
clock coincides exactly with the rate of the laboratory clock which
ticks identically everywhere. Therefore, both the clocks measure
equal time intervals.

\section{Forward-trip position meter}
\begin{figure}[h]
\includegraphics[scale=0.6]{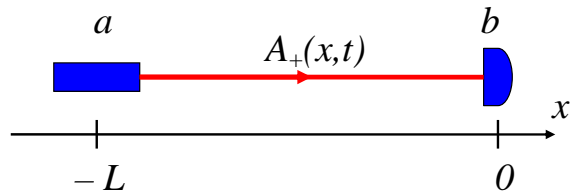}
\caption{A forward-trip position meter. Initially at the state of
rest the built-in clocks in both the laser (test mass $a$) and
detector (test mass $b$) are assumed to be perfectly synchronized.
Detector measures the phase shift of light wave emitted by laser. In
the state of rest the distance between the test masses equals to
$L$.}\label{pic_laser-detector}
\end{figure}

Let us now analyze the operation of the forward trip position meter
illustrated in Fig. \ref{pic_laser-detector}. The system under
consideration consists of only two test masses: laser (test mass
$a$) and detector (test mass $b$) separated by a distance $L$ in the
state of rest. Both the test masses have the built-in clocks, which
initially (at the state of rest) are assumed to be perfectly
synchronized. Laser emits the light wave and detector measures the
phase shift with respect to its clock. Such a position meter (with
slight modifications) might be a constituent part of Mach-Zehnder
interferometer. Similarly to the previous section we will perform
the analysis in both the laboratory and detector frames and compare
the results.

\subsection{Analysis in the inertial frame}
Let the coordinates of the test masses be $x_a(t)=-L+\xi_a(t)$ and
$x_b(t)=0+\xi_b(t)$. Similarly to the previous section we write the
wave emitted by laser in the following form (see Appendix
\ref{sec_optical_noise} for the account of optical noise):
\begin{equation*}
    A_+(x,t)=A_{+0}\exp\left\{-i\omega_0\left[t-\frac{x-x_a(t-(x+L)/c)}{c}\right]\right\}.
\end{equation*}

Detection of the wave and measurement of its phase takes place at
$x=x_b(t)=\xi_b(t)$:
\begin{equation*}
    A_+(x_b(t),t)=A_{+0}\exp\left\{-i\omega_0\left[t-\frac{x_b(t)-
    x_a(t-\tau)}{c}\right]\right\}.
\end{equation*}
Obviously, total variation of the wave equals to
\begin{equation*}
    \delta a(t)=A_{+0}e^{i\omega_0\tau}ik_0\Bigl[\xi_b(t)-\xi_a(t-\tau)\Bigr].
\end{equation*}

The phase shift due to the test masses motion is
\begin{equation}
    \delta\Psi(t)=k_0\Bigl[\xi_b(t)-\xi_a(t-\tau)\Bigr].
    \label{unmeas_phase_shift}
\end{equation}
Let the system move as the entire body, $\xi_a(t)=\xi_b(t)$. It is
interesting that the first non-vanishing term in the expansion of
$\delta\Psi(t)$ into series of $\tau$ is $k_0\dot{\xi}_b(t)\tau$,
i.e. is proportional to the instantaneous velocity of the body. This
does not mean, however, that one is able to measure the latter,
contradicting the relativity principle. This result simply tells us
that the phase shift (\ref{unmeas_phase_shift}) is unmeasurable by
detector. In order to make it measurable we should perform a
coordinate transformation that brings us from the laboratory frame
to the frame of detector --- the inverse transformation of Eqs.
(\ref{coord trans}). However, from the logical point of view it is
more convenient to perform the analysis completely in the frame of
detector.

\subsection{Analysis in the non-inertial frame}
In the reference frame of detector (test mass $b$) coordinates of
the test masses are $x'_a(t)=-L+\delta x'_a(t)$ and $x'_b(t)=0$.

From the viewpoint of detector laser emits electromagnetic wave
described by the following vector-potential:
\begin{align*}
    &A'_+(x',t')\\
    &=A_{+0}\Bigl[1+w'_+(x',t')-w'_+(-L,t'-(x'+L)/c)\Bigr]\\
    &\quad\times\exp\left\{-i\omega_0\left[t'-\frac{x'-x'_a(t'-(x'+L)/c)}{c}\right]\right\}.
\end{align*}
This corresponds to the boundary condition which states that at the
point of laser location $A_+(x'_a(t),t')=A_{+0}e^{-i\omega_0t'}$. In
other words, in the immediate vicinity of the laser light wave
acquires neither localized phase shift due to laser motion nor the
distributed phase shift due to the acceleration of the observer.

At the detector location:
\begin{align*}
    A_+(x'_b(t),t)&=A_{+0}\Bigl[1+w'_+(0,t)-w'_+(-L,t-\tau)\Bigr]\nonumber\\
    &\quad\times\exp\left\{-i\omega_0\left[t'-\frac{x'_b(t)-x'_a(t-\tau)}{c}\right]\right\}.
\end{align*}
Since $x'_b(t)=0$ and $w'_+(0,t)=0$, total variation of the optical
field equals to:
\begin{equation*}
    \delta
    a'(t)=-A_{+0}e^{i\omega_0\tau}\Bigl[ik_0\delta x'_a(t-\tau)+w'_+(-L,t-\tau)\Bigr].
\end{equation*}

The phase shift we are interested in equals to:
\begin{align*}
    &i\,\delta\Psi'(t)=-ik_0\delta x'_a(t-\tau)-w'_+(-L,t-\tau)\\
    &=-ik_0\Bigl[\delta x'_a(t-\tau)+\dot{\xi}_b(t-\tau)\tau+\xi_b(t-\tau)-\xi_b(t)\Bigr].
\end{align*}
If $\delta x_a'(t)=0$ we obtain: $\delta\Psi'(t)\approx
k_0\ddot{\xi}_b(t)\tau^2/2$. This result can be qualitatively
explained in the following way. If $\ddot{\xi}_b(t)>0$ then the
photon moves against the direction of effective gravitational field
with acceleration of the free fall
$g_{\textrm{eff}}=-\ddot{\xi}_b(t)<0$. Thus, photon velocity is
effectively reduced and the optical length increases from $L$ to
$L+|\ddot{\xi}_b(t)\tau^2/2|$. If $\ddot{\xi}_b(t)<0$ than the
photon velocity is increased and optical length is reduced to
$L-|\ddot{\xi}_b(t)\tau^2/2|$.

Substituting into the obtained phase shift solution
(\ref{1st_order_law_of_mot}) for $\delta x'_a(t)=\xi_a(t)-\xi_b(t)$,
we obtain:
\begin{equation}
    \delta\Psi'(t)=k_0\Bigl[\xi_b(t)-\xi_a(t-\tau)-\dot{\xi}_b(t-\tau)\tau\Bigr].
    \label{meas_phase_shift}
\end{equation}
Comparing this result with formula (\ref{unmeas_phase_shift}) we
conclude that the last term in square brackets,
$-\dot{\xi}_b(t-\tau)\tau$, recovers an agreement with the
relativity principle. This term describes the difference in the
proper rate of clock in laser at $x=-L$ and detector at $x=0$. From
the viewpoint of laboratory observer both clocks tick identically.

\subsection{Physical reason of discrepancy between the reference frames}
It is useful to consider the physical reason underlying the
discrepancy of results obtained in the laboratory frame and the
proper frame of detector.

Consideration in the laboratory frame implies that all the clocks
built in the test masses are synchronized with the laboratory clock.
In turn, this requires the knowledge of the velocities of the test
masses with respect to the laboratory frame. Evidently, an observer,
staying in rest in the laboratory, is able to measure the velocities
of the test masses with respect to the latter. In fact, one may
straightforwardly derive from special relativity that phase shift
(\ref{unmeas_phase_shift}) is the one measured by the laboratory
observer who, therefore, is able to measure the common speed of the
test masses.

On the other hand, there is no any ``external observer'' in a system
of $N$ test masses. In such a system test masses are able to
synchronize their clocks only with respect to each other. Evidently,
the absence of knowledge of the test masses velocities with respect
to the laboratory will result in the accuracies of the order of
$(v/c)L$ in clock synchronization. The latter will inevitably enter
the phase shift as $k_0(v/c)L$ which is exactly the term missing in
(\ref{unmeas_phase_shift}) and present in (\ref{meas_phase_shift}).
Therefore, we conclude that the discrepancy between two reference
frames lies in different procedures of clocks synchronization.

From these reasonings it is also clear why calculations of the
round-trip scheme in both the frames produce identical results.

\section{Space-time of an accelerated observer with account for gravitational waves}
The performed analysis can be straightforwardly generalized to take
into account the action of the GWs on position meters. According to
Refs. \cite{1978_inert_grav_eff,1994_fermi} inertial and
gravitational effects do not couple in the first order. Therefore,
one may ``linearly combine'' results of this paper with results of
Ref. \cite{2007_GW_FP_LL} to calculate the response of position
meter to GWs in the proper reference frame of detector. For
instance, consider space-time metric
\begin{align*}
    ds^2=&-(c\,dt')^2\left[1+\frac{2}{c^2}\,a_x(t')x'\right]+
    dx'^2+dy'^2+dz'^2\nonumber\\
    &+\,\frac{1}{2}\,\frac{x'^2-y'^2}{c^2}\,\ddot{h}(t'-z'/c)\,(c\,dt'-dz')^2,
\end{align*}
corresponding to the observer moving with non-geodesic acceleration
$a_x(t')$ along the $x$-axis in the field of weak plane
gravitational wave $h(t'-z'/c)$ propagating along the $z$-axis
normal to the $xy$-plane. In this section we will call the proper
reference frame of an observer the local Lorentz (LL) gauge, since
at $x'=y'=0$ metric is locally flat. Transverse-traceless (TT) gauge
corresponds then to the laboratory frame (see below).

If the coordinate of the test mass relative to an observer equals to
$X_0$ on average (in the state of rest) then the former moves
according to the motion law
\begin{equation*}
    \delta x'(t)=\frac{1}{2}\,X_0h(t)+\xi(t)-\xi_{\textrm{ref}}(t),
\end{equation*}
where $\ddot{\xi}_{\textrm{ref}}(t)=a_x(t)$. Remind, that the
difference between laboratory time and observer time leads to the
negligible 2nd order effects.

Vector-potential of the electromagnetic wave propagating in this
space-time along the $x$-axis can be written in the following form:
\begin{equation*}
    A'_\pm(x',t')=A_{\pm0}\Bigl[1+g'_{\pm}(x',t')+w'_{\pm}(x',t')\Bigr]e^{-i(\omega_0t'\mp k_0x')},
\end{equation*}
where
\begin{align*}
    g'_\pm&(x',t')\approx g'_\pm(x,t)\\
    &=ik_0\left[\frac{1}{4}\,x\dot{h}(t)\,\frac{x}{c}\mp
    \frac{1}{2}\,xh(t)+\frac{c}{2}\int_{t\mp x/c}^{t}h(t_1)dt_1\right],\\
    w'_\pm&(x',t')\approx w'_\pm(x,t)\\
    &=ik_0\Bigl[-\dot{\xi}_{\textrm{ref}}(t)\,\frac{x}{c}\pm
    \xi_{\textrm{ref}}(t)\mp\xi_{\textrm{ref}}(t\mp x/c)\Bigr].
\end{align*}

If one considers, for instance, the response of a round-trip
position meter to GW and fluctuative motions of the test masses
using the method developed in Sec. \ref{sec_round_trip_detector},
the obtained phase shift will be:
\begin{align*}
    &\delta\Psi^{\textrm{TT}}_{\textrm{r.t.}}(t)=\delta\Psi'^{\textrm{LL}}_{\textrm{r.t.}}(t)\\
    &=k_0\Bigl[2\xi_b(t-\tau)-\xi_a(t)-\xi_a(t-2\tau)\Bigr]
    +\frac{\omega_0}{2}\int_{t-2\tau}^{t}h(t_1)dt_1.
\end{align*}
Note that this result can be derived in both the TT and LL gauges.
For small enough $\tau$ ($\Omega\tau\ll1$ in spectral domain)
$\delta\Psi_{\textrm{r.t.}}\approx 2k_0(Lh/2+\xi_b-\xi_a)$. This is
the common result for the LIGO-type GW detectors, where
$\xi_{a,b}(t)$ are the fluctuative displacements of the test masses
that mimic the GW signal $h(t)$.

For a forward-trip coordinate meter corresponding phase shift,
calculated in the LL gauge, will be:
\begin{align}
    \delta\Psi'^{\textrm{LL}}_{\textrm{f.t.}}(t)&=
    k_0\Bigl[\xi_b(t)-\xi_a(t-\tau)-\dot{\xi}_b(t-\tau)\tau\Bigr]\nonumber\\
    &\quad-\frac{1}{4}k_0L\dot{h}(t-\tau)\tau+\frac{\omega_0}{2}\int_{t-\tau}^{t}h(t_1)dt_1.
    \label{LL_forward_trip_phase}
\end{align}
Expanding into series of $\tau$ we obtain
$\delta\Psi_{\textrm{f.t.}}\approx k_0(Lh/2+\xi_b-\xi_a)$ which is
exactly the half of the round-trip phase.

Note that the phase shift (\ref{LL_forward_trip_phase}) differs from
the one that could be obtained in the TT gauge:
\begin{equation*}
    \delta\Psi_{\textrm{f.t.}}^{\textrm{TT}}(t)=
    k_0\Bigl[\xi_b(t)-\xi_a(t-\tau)\Bigr]+\frac{\omega_0}{2}\int_{t-\tau}^{t}h(t_1)dt_1.
\end{equation*}
Comparing this phase shift with the phase shift
(\ref{unmeas_phase_shift}) we may conclude that the TT gauge in GW
physics plays the similar role to the laboratory frame in the
globally flat space-time. For instance, since $g_{00}(x,t)\equiv -1$
in the TT gauge
\cite{1973_gravitation,2003_clas_phys_GW,2005_basics_of_gw_theory},
clock tick identically everywhere. GW manifests itself as the
effective time-dependent optical refraction index; test masses stay
in rest in this gauge. The proper reference frame of detector
corresponds then to the LL gauge, where rate of the clock coincides
with the TT-clock only at the coordinate origin, and GW manifests
itself as the tidal force-field acting on the test masses;
electromagnetic wave is affected only slightly. Therefore, (in full
similarity with Newtonian physics) it is natural that the results
obtained in different gauges do not coincide in accordance with
different procedures of clocks synchronization associated with them
as described in the previous Section.

It will be also interesting to examine how additional term,
$-k_0L\dot{h}(t-\tau)\tau/4$, in Eq. (\ref{LL_forward_trip_phase})
influences the responses of Mach-Zehnder or LISA-type
\cite{1998_LISA} interferometers. This problem requires additional
detailed analysis and we do not consider it in this paper.

The related problem is the transformation of results between
different proper reference frames when analyzing an array of
emitters and receivers. In general this should be performed by the
coordinate transformation from one proper frame to another. However,
such a transformation results in much more cumbersome calculations
than performing the analysis for another proper frame from the
beginning.

\section{Conclusion}
In this paper we developed a method of analyzing the operation of
the optical position meters in the reference frames of their
detectors, which are non-inertial in general. First we studied the
motion of the test masses and propagation of electromagnetic waves
in the space-time of accelerated observer. Then we considered the
operation of the round-trip position meter and found that the phase
shift of light wave calculated in the laboratory (globally inertial)
frame equals to the one calculated in the proper (non-inertial)
reference frame of detector. This coincidence owes to the particular
geometry of the round-trip scheme: phases of both the emitted and
detected light waves are measured at the same spacial point by one
clock. However, for the forward-trip position meter situation is
completely different: the rate of laser and detector clocks,
separated by a large distance, differ greatly. This results in
different phase shifts calculated in the laboratory frame and the
frame of detector. Namely, the former one contradicts the relativity
principle and thus is unmeasurable.

We also discussed the generalization of the developed method to take
into account the action of GWs. We demonstrated that the responses
of the round-trip position meter, calculated in the TT and LL gauges
coincide, while the ones of the forward-trip position meter differ,
in full similarity with the laboratory and detector frames. The
performed analysis could be useful in consideration of the various
types of displacement-noise-free GW detectors, where fluctuative
motion of detector may play the crucial role.

\acknowledgements The author would like to thank S.P. Vyatchanin,
F.Ya. Khalili and K. Somiya for valuable critical remarks on the
paper.

This work was supported by LIGO team from Caltech and in part by NSF
and Caltech grant PHY-0353775 and by Grant of President of Russian
Federation NS-5178.2006.2.

\appendix
\section{Solution of the wave equation}\label{sec_solution_wave_eq}
In this Appendix we solve the 1st order wave equation
(\ref{1st_order_wave_eq}). For briefness we omit all the primes
here:
\begin{align*}
    \frac{1}{c^2}\frac{\partial^2A_\pm^{(1)}}{\partial t^2}-
    \frac{\partial^2A_\pm^{(1)}}{\partial x^2}&=
    \frac{ax}{c^2}\left(\frac{1}{c^2}\frac{\partial^2A_\pm^{(0)}}{\partial t^2}+
    \frac{\partial^2A_\pm^{(0)}}{\partial x^2}\right)\nonumber\\&\quad-
    \frac{\dot{a}x}{c^3}\,\frac{1}{c}\frac{\partial A_\pm^{(0)}}{\partial t}+
    \frac{a}{c^2}\,\frac{\partial A_\pm^{(0)}}{\partial x}.
\end{align*}
Remind, that the 0th order solution is given by formula
(\ref{0th_order_solution}) and we omit the ``c.c.'' terms.

It is convenient to solve this equation in spectral domain. Applying
the theorem of convolution to the right side of the equation we
obtain the 1st order equation in spectral domain:
\begin{multline*}
    -\frac{\Omega^2}{c^2}\,A_\pm^{(1)}(x,\Omega)-
    \frac{\partial^2A_\pm^{(1)}(x,\Omega)}{\partial x^2}=\\
    =-\frac{k_0A_{\pm0}e^{\pm ik_0x}}{c^2}\,a(\Omega-\omega_0)
    \left[2k_0x-\frac{\Omega-\omega_0}{c}\,x\mp i\right].
\end{multline*}
Let us introduce the following notations:
\begin{equation*}
    A^{(1)}_\pm(x,\Omega)=A_\pm(x),\quad
    \frac{\Omega}{c}=k,\quad
    \frac{A_{\pm0}a(\Omega-\omega_0)}{c^2}=B.
\end{equation*}
In this notation equation takes the following form:
\begin{equation*}
    \frac{d^2A_\pm}{dx^2}+k^2A_\pm=Be^{\pm ik_0x}(3k_0^2x-k_0kx\mp ik_0).
\end{equation*}

We solve this equation with the method of variation of constants
(see any ODE handbook):
$A_\pm(x)=C_{\pm1}(x)e^{ik_0x}+C_{\pm2}(x)e^{-ik_0x}$. The set of
equations for $C_{\pm1,2}$ is:
\begin{align*}
    &\frac{dC_{\pm1}}{dx}e^{ikx}+\frac{dC_{\pm2}}{dx}e^{-ikx}=0,\\
    &ik\frac{dC_{\pm1}}{dx}e^{ikx}-ik\frac{dC_{\pm2}}{dx}e^{-ikx}=Be^{\pm
    ik_0x}(3k_0^2x-k_0kx\mp ik_0).
\end{align*}
Straightforward integration leads to:
\begin{widetext}
\begin{align*}
    C_{+1}(x)&=C_{+10}+\frac{B}{2k(k_0-k)^2}\,e^{i(k_0-k)x}
    \Bigl[x(-3k_0^2+4k_0^2k-k_0k^2)-2ik_0^2\Bigr],\\
    C_{+2}(x)&=C_{+20}+\frac{B}{2k(k_0+k)^2}\,e^{i(k_0+k)x}
    \Bigl[x(3k_0^2+2k_0^2k-k_0k^2)+(2ik_0^2-2ik_0k)\Bigr],\\
    C_{-1}(x)&=C_{-10}+\frac{B}{2k(k_0+k)^2}\,e^{-i(k_0+k)x}
    \Bigl[x(3k_0^2+2k_0^2k-k_0k^2)+(-2ik_0^2+2ik_0k)\Bigr],\\
    C_{-2}(x)&=C_{-20}+\frac{B}{2k(k_0-k)^2}\,e^{-i(k_0-k)x}
    \Bigl[x(-3k_0^2+4k_0^2k-k_0k^2)+2ik_0^2\Bigr].
\end{align*}
\end{widetext}
Constants of integration are derived from the boundary condition
(\ref{sum}, \ref{initial_conditions}):
\begin{align*}
    A^{(1)}(x,\Omega)&=A_+^{(1)}(x,\Omega)+A_-^{(1)}(x,\Omega),\\
    A_+^{(1)}(0,\Omega)&=A_-^{(1)}(0,\Omega)=0.
\end{align*}

Therefore, we obtain the following solution of the 1st order in
spectral domain:
\begin{align*}
    A_\pm^{(1)}(x,\Omega)&=\frac{1}{c^2}\,A_{\pm0}e^{\pm
    ik_0x}a(\Omega-\omega_0)\\
    &\quad\times\biggl\{\eta(\Omega)x\pm
    ic\zeta(\Omega)\Bigl[1-e^{(\Omega/c-k_0)x}\Bigr]\biggr\},
\end{align*}
where
\begin{align*}
    \eta(\Omega)&=\frac{-3\omega_0^4+\omega_0^3\Omega+3\omega_0^2\Omega^2-\omega_0\Omega^3}
    {(\omega_0-\Omega)^2(\omega_0+\Omega)^2},\\
    \zeta(\Omega)&=\frac{-5\omega_0^3+2\omega_0^2\Omega-\omega_0\Omega}
    {(\omega_0-\Omega)^2(\omega_0+\Omega)^2}.
\end{align*}
Let us introduce the notation:
\begin{align*}
    w_\pm(x,\Omega)&=\frac{1}{c^2}\,a(\Omega-\omega_0)\\
    &\quad\times\biggl\{\eta(\Omega)x\pm
    ic\zeta(\Omega)\Bigl[1-e^{\pm(\Omega/c-k_0)x}\Bigr]\biggr\}.
\end{align*}
Then
\begin{multline*}
    A^{(1)}_\pm(x,t)=\int_{-\infty}^{+\infty}A_{\pm0}w_\pm(x,\Omega)
    e^{\pm ik_0x}e^{-i\Omega t}\,\frac{d\Omega}{2\pi}\\
    =A_{\pm0}e^{-i(\omega_0t\mp k_0x)}\int_{-\infty}^{+\infty}w_\pm(x,\Omega+\omega_0)
    e^{-i\Omega t}\,\frac{d\Omega}{2\pi}.
\end{multline*}
It is straightforward to verify that in the $\Omega\ll\omega_0$
limit
\begin{equation*}
    \eta(\Omega+\omega_0)\approx\frac{\omega_0}{\Omega},\qquad
    \zeta(\Omega+\omega_0)\approx-\frac{\omega_0}{\Omega^2},
\end{equation*}
and
\begin{equation*}
    w_\pm(x,\Omega+\omega_0)=\frac{a(\Omega)}{\Omega^2}\biggl[\frac{\Omega}{c}\,k_0x\mp
    ik_0\Bigl(1-e^{\pm i\Omega x/c}\Bigr)\biggr].
\end{equation*}
Remind now, that $a(\Omega)/\Omega^2=-\xi_{\textrm{ref}}(\Omega)$.
Ultimately, in time domain we obtain the following solution:
\begin{align*}
    A^{(1)}_\pm(x,t)&=A_{\pm0}w_\pm(x,t)e^{-i(\omega_0t\mp k_0x)},\\
    w_\pm(x,t)&=\int_{-\infty}^{+\infty}w_\pm(x,\Omega+\omega_0)e^{-i\Omega t}\,
    \frac{d\Omega}{2\pi}\\
    &=-ik_0\dot{\xi}_{\textrm{ref}}(t)\,\frac{x}{c}\pm
    ik_0\Bigl[\xi_{\textrm{ref}}(t)-\xi_{\textrm{ref}}(t\mp x/c)\Bigr].\nonumber
\end{align*}
Remind, that the prime should be inserted everywhere, since we work
in the non-inertial frame.

\section{Influence of the optical noise}\label{sec_optical_noise}
In this Appendix we briefly consider the influence of laser optical
noise on the responses of round- and forward-trip position meters.

In general optical noise can be taken into account by adding the
following term to initial vector-potentials \cite{2007_GW_FP_LL}:
\begin{equation*}
    a_+(x,t)=\int_{-\infty}^{+\infty}a_+(\omega_0+\Omega)
    e^{-i\Omega\left(t-\frac{x-x_0}{c}\right)},
\end{equation*}
where $x_0$ is the reference point for specific problem. For
instance, for the round-trip meter $x_0=0$ and for the forward-trip
meter $x_0=-L$. Since optical noise, in practice, is comparable to
other noises in their magnitude, one can neglect the interaction
between GW and acceleration fields with it.

Now substituting optical noise into the corresponding boundary
problems we obtain in spectral domain:
\begin{align*}
    \delta a(\omega_0+\Omega)=&-a_+(\omega_0+\Omega)e^{2i(\omega_0+\Omega)\tau}\\
    &+A_{+0}e^{2i\omega_0\tau}ik_0\Bigl[\xi_a-2\xi_be^{i\Omega\tau}+\xi_ae^{2i\Omega\tau}\Bigr],
\end{align*}
for the round-trip meter in both the laboratory and proper frame of
detector;
\begin{align*}
    \delta a(\omega_0+\Omega)&=a_+(\omega_0+\Omega)e^{i(\omega_0+\Omega)\tau}\\
    &\quad+A_{+0}e^{i\omega_0\tau}ik_0\Bigl[\xi_b-\xi_ae^{i\Omega\tau}\Bigr],
\end{align*}
for the forward-trip meter in the laboratory frame and
\begin{align*}
    \delta a(\omega_0+\Omega)&=a_+(\omega_0+\Omega)e^{i(\omega_0+\Omega)\tau}\\
    &\quad+A_{+0}e^{i\omega_0\tau}ik_0\Bigl[\xi_b-\xi_ae^{i\Omega\tau}+
    i\Omega\tau\xi_be^{i\Omega\tau}\Bigr],
\end{align*}
for the forward-trip meter in the proper frame of detector. Here
$\xi_{a,b}=\xi_{a,b}(\Omega)$. GW can be taken into account
straightforwardly.


\end{document}